\documentclass[twocolumn,showpacs,amsmath,amssymb]{revtex4}

\usepackage{dcolumn}
\usepackage{bm}
\usepackage{epsfig}
\usepackage{amsfonts}

\begin{document}


\title{Could dark energy be measured in the lab?}

\author{Christian Beck}
\affiliation{
School of Mathematical Sciences \\ Queen Mary, University of
London \\ Mile End Road, London E1 4NS, UK}
\email{c.beck@qmul.ac.uk}
\homepage{http://www.maths.qmul.ac.uk/~beck}
\author{Michael C. Mackey}
\affiliation{Centre for Nonlinear Dynamics in Physiology and
Medicine \\ Departments of Physiology, Physics and Mathematics \\
McGill University, Montreal, Quebec, Canada}
\email{mackey@cnd.mcgill.ca}
\homepage{http://www.cnd.mcgill.ca/people_mackey.html}
\altaffiliation{ also: Mathematical Institute, University of
Oxford, 24-29 St Giles', Oxford OX1 3LB, UK}

\date{\today}

\vspace{2cm}

\begin{abstract}
The experimentally measured spectral density of current noise in
Josephson junctions provides direct evidence for the existence of
zero-point fluctuations. Assuming that the total vacuum energy
associated with these fluctuations cannot exceed the presently
measured dark energy of the universe, we predict an upper cutoff
frequency of $\nu_c=(1.69\pm 0.05) \times 10^{12}$ Hz for the
measured frequency spectrum of zero-point fluctuations in the
Josephson junction. The largest frequencies that have been reached
in the experiments are of the same order of magnitude as $\nu_c$
and provide a lower bound on the dark energy density of the
universe.
It is shown that suppressed zero-point fluctuations
above a given cutoff frequency can lead to $1/f$ noise.
We propose an experiment which may help to measure
some of the properties of dark energy in the lab.

\end{abstract}


\date{\today}          
\maketitle


\section{Introduction}

In his ``second theory" of black-body radiation, Planck
\cite{planck14} (cf. also \cite{planck88}) found the average
energy of a collection of  oscillators at temperature $T$ and
frequency $\nu$ to be
\begin{equation}
    {\bar U}(\nu,T) = \frac{1}{2} h \nu  + \frac{h \nu }{\exp(h \nu /kT)-1}.
     \label{energy}
\end{equation}
The first (temperature independent) term  is now referred to as
the zero-point energy and commonly related to vacuum fluctuations.
The second term gives rise \cite{planck14,nernst16}  to the Planck
black body spectrum
\begin{equation}
    \rho(\nu,T) = \frac{8 \pi h \nu^3}{c^3}
    \frac{1}{\exp(h \nu /kT)-1}
     \label{spectrum}
\end{equation}
that is relatively flat for $h \nu << kT$ and which approaches
zero for $h \nu >> kT$.

In spite of early convictions by  some investigators that the 
zero-point energy term in Equation (\ref{energy}) would not have any
experimental correlate, this has not been the case. Indeed, the
zero-point term has proved important in explaining X-ray
scattering in solids \cite{debye14}; understanding of the Lamb
shift between the s and p levels in hydrogen
\cite{welton48,power66}; predicting the Casimir effect
\cite{casimir48a,casimir48b,bressi}; understanding the origin of Van der
Waals forces
\cite{casimir48a};
interpretation of the Aharonov-Bohm effect
\cite{boyer73a,boyer87}; explaining Compton scattering
\cite{welton48}; and predicting the spectrum of noise in
electrical circuits \cite{koch80,koch82,weber53,senitzky60}. It is
this latter effect that concerns us here.

Koch et al. \cite{koch82} measured the frequency spectrum of
current fluctuations in Josephson junctions. At low temperatures
and high frequencies the experimental spectrum is dominated by
zero-point fluctuations, confirming  the physical relevance of the
zero-point term in Equation (\ref{energy}) up to frequencies of
the order $\nu_{max}=6\times 10^{11}$ Hz. Here we re-analyze their
experimental results in light of  recent astronomical estimates of
dark energy density in the universe
\cite{bennett03,spergel03,peebles,beck04}.

Our hypothesis  is that the signature of zero-point fluctuations
measured by Koch et al. imply a non-vanishing vacuum energy
density in the universe. This vacuum energy  would have large
scale gravitational effects, and cannot exceed the measured dark
energy density of the universe as determined in astronomical
measurements \cite{bennett03,spergel03}. On this basis we predict
a cutoff frequency ($\nu_c$) for the zero-point fluctuations in
Josephson junction experiments, which is only slightly larger than
the maximum frequency $\nu_{max}$ reached in Koch et al.'s 1982
experiment. Future experiments, based on
Josephson junctions that operate in the THz region \cite{wang,divin},
could thus help to clarify whether
this cutoff exists and whether the dark energy of the universe is
related to the vacuum fluctuations that play a role in the
Josephson junction experiments.

\section{Estimating a cutoff frequency for
zero-point fluctuations}

%

If Planck \cite{planck14} and Nernst \cite{nernst16}  had used the
relation $\rho(\nu,T) =  {8 \pi h \nu^2}\bar U(\nu,T)/{c^3} $,
then instead of Equation (\ref{spectrum}) they would have obtained
\begin{eqnarray}
    \rho(\nu,T) &=& \frac{8 \pi \nu^2}{c^3}
    \left [ \frac{1}{2}h \nu  + \frac {h \nu }{\exp(h \nu /kT)-1} \right ] \nonumber \\
    &=& \frac{4   \pi  h \nu^3}{c^3}
    \left [ 1  + \frac {2 }{\exp(h \nu /kT)-1} \right ] \nonumber \\
    &=& \frac{4   \pi  h \nu^3}{c^3}
    \coth \left ( \frac{h\nu}{kT} \right ).
     \label{nu-spectrum}
\end{eqnarray}
Equation (\ref{nu-spectrum}), which is correct from the
perspective of quantum electrodynamics \cite{milloni94}, predicts
that if all frequencies $\nu$  are taken into account then there
should be an infinite energy per unit volume since
\begin{displaymath}
    \lim_{\nu_c \to \infty} \int_0^{\nu_c}  \rho(\nu,T) d\nu
    \nonumber
\end{displaymath}
diverges. To avoid this one could  introduce a  cutoff frequency
$\nu_c<\infty$.

Split the total energy density into
\begin{equation}
\rho(\nu,T)=\rho_{vac}(\nu) +\rho_{rad}(\nu,T), \label{4}
\end{equation}
where
\begin{equation}
\rho_{vac}(\nu)=\frac{4\pi h\nu^3}{c^3} \label{vac}
\end{equation}
is due to zero-point fluctuations, and
\begin{equation}
\rho_{rad}(\nu , T)=\frac{8\pi h \nu^3}{c^3}
\frac{1}{\exp{(h\nu/kT)}-1}. \label{rad}
\end{equation}
corresponds to the radiation energy density generated by photons
of energy $h\nu$. Integration of (\ref{vac}) up to
$\nu_c$ yields
\begin{equation}
     \int_0^{\nu_c}  \rho_{vac}(\nu) d\nu =
    \frac {4   \pi  h}{c^3} \int_0^{\nu_c} \nu^3
    d\nu = \frac {   \pi  h}{c^3}  \nu_c^4,
    \label{total-2}
\end{equation}
while integration of (\ref{rad}) over all frequencies
yields the well-known Stefan-Boltzmann law
\begin{equation}
\int_0^\infty \rho_{rad}(\nu,T)d\nu =
\frac{\pi^2k^4}{15\hbar^3c^3} T^4.
\end{equation}
Suppose Equation (\ref{vac}) is 
valid only up to a cutoff frequency $\nu_c$,
due to new but as yet unknown physics. 
How might we determine $\nu_c$? We propose using  estimates of the
dark energy density to place an upper limit on the value
calculated from Equation (\ref{total-2}).

Current estimates \cite{bennett03,spergel03} indicate
that dark energy constitutes $73\%$ of all energy in the universe.
To calculate the dark energy density $\rho_{dark}$ we need the
critical energy density $\rho_c$ of a flat universe (the  data of
\cite{spergel03} indicate that the universe is flat), which is
$\rho_c = 10.539 h_{\mbox{Hubble}}^2 \;\mbox{GeV/m}^3 = 10.539
\times (0.71\pm 0.04)^2 \;\mbox{GeV/m}^3$.  Finally, we have
\begin{equation}
    \rho_{dark} = 0.73 \rho_c  = (3.9 \pm 0.4) \quad \mbox{GeV/m}^3
\end{equation}
If we set
\begin{equation}
    \frac{   \pi  h}{c^3}  \nu_c^4 \simeq \rho_{dark}
\end{equation}
then
\begin{equation}
    \nu_c \simeq (1.69 \pm 0.05) \times 10^{12} \quad \mbox{Hz}.
\label{cutoff}
\end{equation}

\section{Measurements of
zero-point fluctuations in Josephson junctions}

The behavior of a resistively shunted Josephson junction is
modeled as a particle that moves in a tilted periodic potential,
and the effect of the noise current is to produce random
fluctuations of the tilt angle \cite{koch80}.  This situation
is captured  by the stochastic differential equation
\begin{equation}
 \frac{\hbar
C}{2e}\ddot{\delta}+\frac{\hbar}{2eR}\dot{\delta} +I_0\sin \delta
=I +I_N.
\end{equation}
 Here $\delta$ is the phase difference across the
junction, $R$ is the shunt resistor, $C$ the capacitance of the
junction, $I$ is the mean current, $I_0$ the noise-free critical
current, and $I_N$ is the noise current. As shown in
\cite{callen51,koch80}, the junction noise current should have a
spectral density given by
\begin{eqnarray}
 S(\nu)&=&\frac{2h\nu}{R}\coth
\left( \frac{h\nu}{kT} \right) \nonumber \\ &=&\frac{4h\nu}{R}
\left( \frac{1}{2}+\frac{1}{\exp (h\nu/kT)-1} \right)  .
\label{power}
\end{eqnarray}

The first term in  Equation  (\ref{power}) is due to vacuum
fluctuations, and the second one is due to ordinary Bose-Einstein
statistics.  This predicted spectral behaviour  has been
experimentally verified in the work of \cite{koch82}
measuring the current noise in a resistively shunted Josephson
junction at two different temperatures.
Further, the computed cutoff frequency (\ref{cutoff}) is less than
one order of magnitude larger than the highest frequency used in
these experiments.
Fig. 1 shows how well the predicted form of the power spectrum
(\ref{power}) is experimentally verified up to frequencies of
order $6\times 10^{11}$ Hz (note that no fitting parameters are
used in this figure). For more recent theoretical work on the quantum
noise theory of Josephson junctions, see \cite{gardiner,levinson}.

\begin{figure}
\epsfig{file=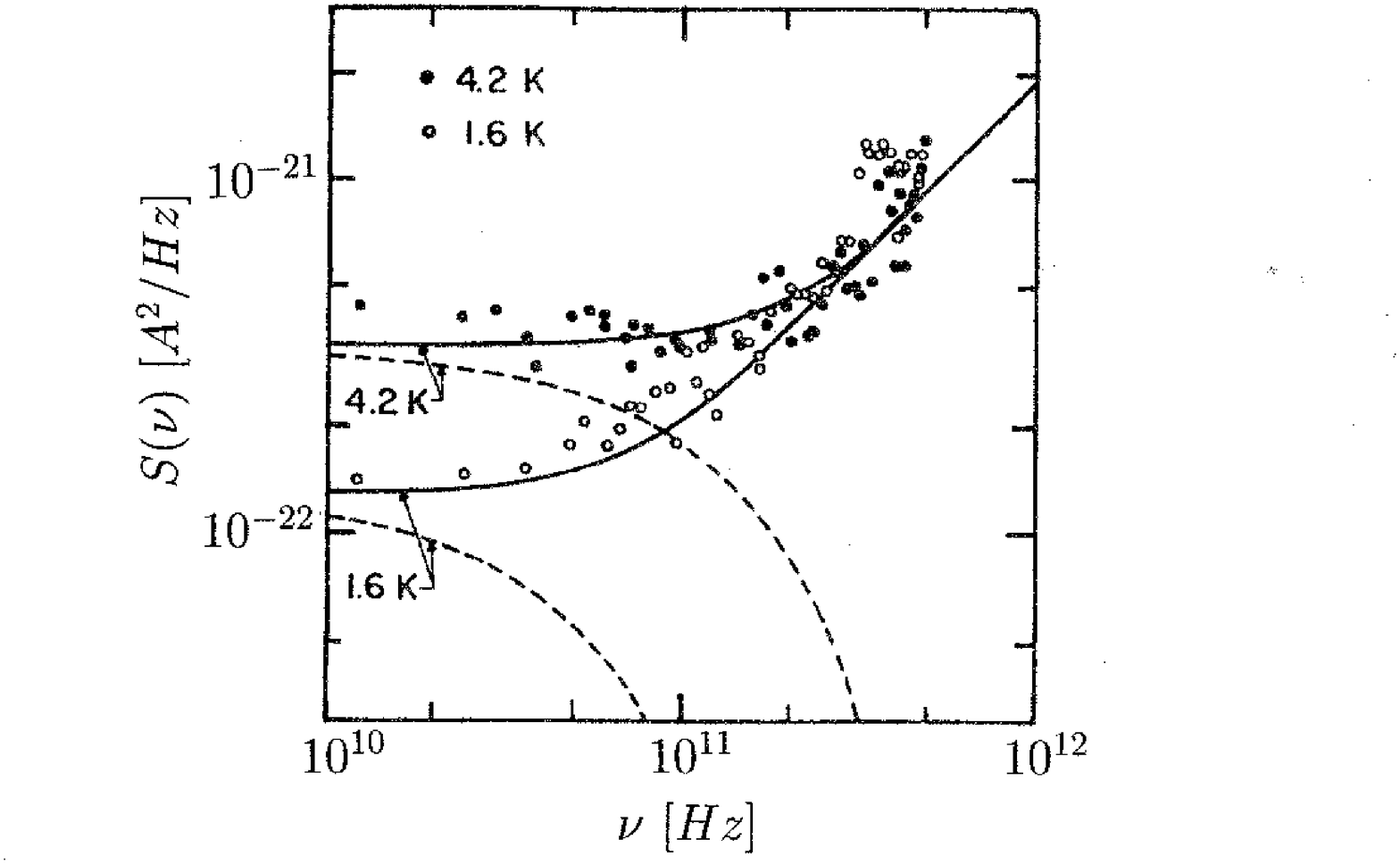, width=9cm, height=6.5cm}
\caption{Spectral density of current noise as measured in Koch et
al.'s experiment \cite{koch82} for two different temperatures. The
solid line is the prediction of Equation (\ref{power}), whereas
the dashed line is given by $(4h\nu/R)(exp(h\nu/kT)-1)^{-1}$.}

\end{figure}

Zero-point fluctuations thus have theoretically predicted and
experimentally measured effects in Josephson junctions.  We
therefore expect that the energy density associated with these
fluctuations has physical meaning as well: It is a prime candidate
for dark energy, being isotropically distributed and temperature
independent. Note that the experimentally measured fluctuations
in Fig.~1 are physical reality and 
have to be distinguished from
'theoretical' zero-point fluctuations that just formally
enter into QED calculations without any cutoff. The vacuum energy
associated with the measured data in Fig.~1 cannot be easily discussed
away.


Assuming that the vacuum energy associated with 
the measured fluctuations in Fig.~1
is physically relevant,
we predict that the measured spectrum in Josephson junction
experiments must exhibit a cutoff at the critical frequency
$\nu_c$.  If not, the corresponding vacuum energy density would
exceed the currently measured dark energy density of the universe.
In future experimental measurements that may reach
higher frequencies one would have to carefully distinguish
between intrinsic cutoffs (due to experimental constraints)
and fundamental cutoffs (due to new physics).

\section{Implications for dark energy from present and future
experiments}

\subsection{Lower bound on dark energy density}

The largest frequency reached in the Koch et al. \cite{koch82}
experiment was $\nu_{max} \simeq 6\times 10^{11}Hz\approx
\frac{1}{3} \nu_c$. From (\ref{total-2}) this implies a minimum
value of dark energy density in the universe:
\begin{equation}
\rho_{dark} \geq \frac{\pi h}{c^3}\nu_{max}^4=0.062 \;
\mbox{GeV}/\mbox{m}^3
\end{equation}
If larger frequencies $\nu_{max}$ could be reached in a similar
experiment, they would provide a better lower bound.

\subsection{$h\nu_c$ and neutrino masses}

The  energy associated with the computed  cutoff frequency $\nu_c$
\begin{equation}
E_c=h\nu_c=(7.00\pm 0.17)\times 10^{-3} eV
\end{equation}
coincides with current experimental estimates of neutrino masses.
The LMA (large-mixing angle) solution of the solar neutrino problem
yields a mass square difference of roughly $\Delta m^2_{sun}\simeq
7\times 10^{-5}\; \mbox{eV}^2$ between two neutrino species
\cite{neutrino}. Assuming a hierarchy of neutrino masses, this
gives a neutrino mass of the order of magnitude $m_\nu\simeq 8\times
10^{-3}\;\mbox{eV}$.

If this coincidence is confirmed in future experiments, one might
try to develop a theory that links the cutoff frequency of the
zero-point fluctuations to an as yet unknown property of the neutrino
sector of the standard model. 
For previous work that relates the dark energy scale to the mass of
neutrinos, see \cite{xyz}.
Generally, in quantum field theory
bosons are associated with positive vacuum energy and fermions
with negative energies \cite{wess}. In supersymmetric models both
contributions cancel exactly. To explain a coincidence of the type
$h\nu_c \simeq m_\nu c^2$, a possible idea would be that negative
vacuum energy associated with neutrinos (or neutrino-like
particles) might cancel positive
vacuum energy associated with photons as soon as the energy
$E=h\nu$ exceeds the neutrino rest mass. A toy model of this type
is worked out in section V.


\subsection{Effective degrees of freedom contributing
to dark energy}


Photons and other particles contribute to the total vacuum energy
density of the universe. General quantum field theoretical
considerations imply that a particle of mass $m$ and spin $j$
makes a contribution\cite{wess}
\begin{equation}
\rho_{vac}=\frac{1}{2}(-1)^{2j}(2j+1)\int \frac{d^3k}{(2\pi)^3}
\sqrt{{\bf k}^2+m^2}
\label{vacontri}
\end{equation}
in units where $\hbar =c=1$. Here ${\bf k}$ represents the
momentum and the energy is given by $ E=\sqrt{{\bf k}^2+m^2}$. The
integral is divergent and the actual contribution depends on the
regularization
scheme chosen.

It is likely that the Josephson junction experiment only measures
vacuum fluctuations that 
couple to electric charge, since this
experiment is purely based on electromagnetic interaction
(see also \cite{kiefer} on possible interactions of
mesoscopic quantum systems with gravity). Thus
this experiment is likely to see only a fraction $\kappa <1$ of
the total dark energy of the universe. This would modify the
expected cutoff frequency as
\begin{equation} \nu_c=\left(
\kappa  \rho_{dark}^{total}\right)^{1/4}\left( \frac{c^3}{\pi h}
\right)^{1/4}.
\end{equation}
In particular, a small $\kappa$ can significantly lower the cutoff
frequency. A measurement of $\kappa$ would thus give information
on the effective number of degrees of freedom that produce the
entire dark energy density of the universe.

\section{Dark energy and {1/f} noise} In the following we
consider a simple model where a bosonic contribution to vacuum
energy is  suppressed by a fermionic contribution as soon as the
energy exceeds $h\nu_c=mc^2$, where $m$ is the mass of the fermion
under consideration.

Assume $j=1/2$. From Equation (\ref{vacontri}) we obtain the fermionic
contribution to the vacuum energy as
\begin{eqnarray}
\rho_{vac}^{ferm}&=&-\int \frac{d^3k}{(2\pi)^3}
\sqrt{{\bf k}^2+m^2} \nonumber \\
&=& -\frac{1}{2\pi^2} \int_0^{k_{max}}k^2\sqrt{k^2+m^2}dk,
\end{eqnarray}
where $k=|{\bf k}|$ and $k_{max}$ is a suitable upper cutoff.
Transforming from  $k$ to  $E=\sqrt{k^2+m^2}$ this can be written
as
\begin{equation}
\rho_{vac}^{ferm}=-\frac{1}{2\pi^2} \int_m^{E_{max}}\sqrt{E^2-m^2}E^2 dE.
\end{equation}
Additionally  the massless boson contributes with
\begin{equation}
\rho_{vac}^{bos}=+\frac{1}{2\pi^2} \int_0^{E_{max}}E^3dE,
\end{equation}
in agreement with Equation (\ref{total-2}), setting $E=h\nu$ and
$\hbar= c=1$. Adding the two contributions, one obtains
\begin{equation}
\rho_{vac}=\frac{1}{2\pi^2} \int_0^{E_{max}} (E^3-\sqrt{E^2-m^2}E^2 \theta
(E-m)) dE \label{20}
\end{equation}
where the $\theta$-function is defined by
\begin{equation}
\theta (x) =\left\{ \begin{array}{ll} 1 & x\geq 0 \\ 0 & x<0.
\end{array} \right.
\end{equation}
The integrand in Equation (\ref{20}), divided by $E^2/\pi^2$,
represents the effective zero-point energy of this problem.
Correlated vacuum fluctuations of this type would thus produce in
Josephson junctions the power spectrum
\begin{equation}
S(\nu)=\frac{4}{R} \left\{
\begin{array}{ll}
\frac{1}{2}h\nu & h\nu \leq mc^2 \\ \frac{1}{2}(h\nu -
\sqrt{h^2\nu^2-m^2c^4}) & h\nu >mc^2.
\end{array}
\right. \label{23}
\end{equation}
There is a rapid decrease of spectral power above  the critical
frequency $h\nu_c=mc^2$. For frequencies $h\nu
>mc^2$ Equation (\ref{23})
implies
\begin{equation}
S(\nu)= \frac{2}{R} h\nu \left( 1-\sqrt{1-\frac{m^2c^4}{h^2\nu^2}}
\right) .
\end{equation}
For large $h\nu$
\begin{equation}
 \sqrt{1-\frac{m^2c^4}{h^2\nu^2}} \approx
1-\frac{m^2c^4}{2h^2\nu^2},
\end{equation}
and we have
\begin{equation}
S(\nu) = \frac{1}{R}m^2c^4 \frac{1}{h\nu}. \label{1f}
\end{equation}
Thus, asymptotically, the vacuum fluctuation spectrum is inversely
proportional to $\nu$ so suppressed vacuum fluctuations produce
$1/f$ noise. $1/f$ noise is commonly observed  in many electric
circuits, and was also observed in Koch et al.'s experiment but
was subtracted from the data \cite{koch82}. Our simple theoretical
considerations show that high-frequency $1/f$ noise 
can arise naturally if bosonic
vacuum fluctuations are suppressed by fermionic ones. If the
coefficient multiplying $1/\nu$ in Equation (\ref{1f}) is measured
in the experiment, then it can be used to determine the cutoff
scale $h\nu_c= mc^2$.

\section{Conclusion}

We propose a repeat of the experiments of Koch et al. with new
generations of Josephson junctions at  higher frequencies. If it
is possible to increase the maximum frequency by a factor of about
3, then this experiment could provide valuable information on the
nature of dark energy. 
If the vacuum energy associated with the fluctuations measured in
Fig.~1 is physically relevant, 
then we predict  that a deviation from linear
growth of $S(\nu)$ will be seen at higher frequencies, and in fact
a rapid decrease of zero-point power near the critical frequency
$\nu_c$ is expected. If this is not seen in the experiment, then
we must conclude that  the dark energy of the universe probably
has nothing
to do with vacuum fluctuations at all but is purely classical.
Alternately, if this decrease is not observed another
interpretation would be that the Josephson junction experiment is
insensitive to the process which cancels the photonic vacuum
energy at large frequencies. If the frequency cutoff is observed,
it could be used to determine the fraction $\kappa$ of dark energy
density that is produced by electromagnetic processes. Moreover,
if the Josephson junction experiment is repeated  at different
temperatures, then a possible temperature dependence of $\nu_c$
could provide information on whether the dark energy density is
really independent of the expansion of the universe (i.e. its
temperature) or whether it changes slightly  with the expansion
(as in the  models \cite{peebles88}). Finally,
we think that it could be interesting to analyze
experimentally observed high-frequency
$1/f$ noise in electrical circuits under
the hypothesis that it could be a possible manifestation of
suppressed zero-point fluctuations.






\begin{thebibliography}{99}
\bibitem{planck14} Planck, M. (1914), {\em The Theory
of Heat Radiation}. P. Blakistons's Son \& Co.
\bibitem{planck88} Planck, M. (1988), {\em The Theory of
Heat Radiation}. Tomash Publishers and American Institute of
Physics.
\bibitem{nernst16} Nernst, W. (1916), Verh. Dtsch. Phys. Ges.,
{\bf 18}, 83
\bibitem{debye14} Debye, P. (1914), Ann d. Phys {\bf 43}, 49
\bibitem{welton48} Welton T. (1948), Phys. Rev. {\bf 74}, 1157
\bibitem{power66} Power, E. (1966), Am. J. Phys. {\bf 34}, 516
\bibitem{casimir48a} Casimir, H. (1948), Kon. Ned. Akad.
Wetensch. {\bf 51B}, 793
\bibitem{casimir48b} Casimir, H. and Polder, D. (1948),
Phys. Rev. {\bf 73}, 360
\bibitem{bressi} Bressi, G. et al. (2002), Phys. Rev. Lett. {\bf 88},
041804
\bibitem{boyer73a} Boyer, T. (1973), Phys. Rev. D {\bf 8}, 1679
\bibitem{boyer87} Boyer, T. (1987), Phys. Rev. A {\bf 36}, 5083
\bibitem{koch80} Koch, R.H., van Harlingen, D. and
Clarke J. (1980), Phys. Rev. Lett. {\bf 45}, 2132
\bibitem{koch82} Koch, R.H., van Harlingen, D. and
Clarke J. (1982), Phys. Rev. B {\bf 26}, 74
\bibitem{weber53} Weber, J. (1953), Phys. Rev. {\bf 90}, 977
\bibitem{senitzky60} Senitzky, I. (1960), Phys. Rev. {\bf 119}, 670
\bibitem{bennett03} Bennett, C.L. et al. (2003),
Astrophys. J. Supp. Series {\bf 148}, 1 (astro-ph/0302207)
\bibitem{spergel03} Spergel D.N. et al. (2003),
Astrophys. J. Supp. Series {\bf 148}, 148 (astro-ph/0302209)
\bibitem{peebles} Peebles, P.J.E. and Ratra, B. (2003),
Rev. Mod. Phys. {\bf 75}, 559
(astro-ph/0207347);
Weinberg, S. (2000), astro-ph/0005265;
Trodden, M. and Caroll, S. (2004), astro-ph/0401547;
Dolgov, A.D. (2004), hep-ph/0405089, Frampton, P.H. (2004), astro-ph/0409166
\bibitem{beck04} Beck, C. (2004) Phys. Rev. D {\bf 69}, 
123515 (astro-ph/0310479)
\bibitem{wang} Wang, H.B., Wu, P.H. and Yamashita, T. (2001),
Phys. Rev. Lett. {\bf 87}, 107002
\bibitem{divin} Divin, Y.Y. et al. (2002), Physica C {\bf 372-376},
416 
\bibitem{milloni94} Milonni, P.W. (1994),
{\em The Quantum Vacuum: An Introduction to Quantum
Electrodynamics.} Academic Press Inc., Boston
\bibitem{callen51} Callen, H. and Welton, T. (1951) Phys. Rev. {\bf 83}, 34
\bibitem{gardiner} Gardiner, C.W. (1991), {\em Quantum Noise},
Springer, Berlin
\bibitem{levinson} Levinson, Y. (2003), Phys. Rev. B {\bf 67},
184505
\bibitem{neutrino} Aliani P. et al. (2003), hep-ph/0309156
\bibitem{xyz} Hung, P.Q. (2000), hep-ph/0010126,
Hung, P.Q. and Paes, H. (2003), astro-ph/0311131, Fardon R., Nelson A.E.
and Weiner, N. (2003), astro-ph/0309800, Kaplan, D.B., Nelson A.E. and
Weiner, N. (2004), hep-ph/0401099
\bibitem{wess} Wess, P. (1990), {\em Introduction to
Supersymmetry and Supergravity}, World Scientific, Singapore
\bibitem{kiefer} Kiefer, C. and Weber, C. (2004), gr-qc/0408010
\bibitem{peebles88} Peebles, P.J.E. and Ratra, B. (1988),
Astrophys. J. {\bf 325}, L17;
Turner, M.S. and White, M. (1997),
Phys. Rev. D {\bf 56}, 4439;
Caldwell, R.R., 
Dave, R. and Steinhardt P.J. (1998),
Phys. Rev. Lett. {\bf 80}, 1582
\end{thebibliography}
\end{document}